# No time loophole in Bell's theorem; the Hess–Philipp model is non-local

R.D. Gill,[*][†] G. Weihs,[‡] A. Zeilinger[§] and M. Żukowski[¶]

**Hess and Philipp recently claim in this journal that proofs of Bell's theorem have overlooked the possibility of time dependence in local hidden variables, hence the theorem has not been proven true. Moreover they present what is claimed to be a local realistic model of the EPR correlations. If this is true then Bell's theorem is not just unproven, but false. We refute both claims. Firstly we explain why time is not an issue in Bell's theorem, and secondly show that their hidden variables model violates Einstein separability. Hess and Philipp have overlooked the freedom of the experimenter to choose settings of a measurement apparatus at will: any setting could be in force during the same time period.**

The recent papers (1, 2, 3) by Hess and Philipp have drawn a lot of attention, especially since (1, 2), in this journal, were featured on *Nature*'s web pages and from there reached the popular press in many countries. It is claimed that Bell (4) overlooked the role of time, and hence his theorem, that quantum mechanics is incompatible with local realism, is unproven. This is supported with an apparent counter-example: a supposedly local hidden variables model for the singlet correlations.

While Hess and Philipp correctly state that time was ignored in the original derivation of Bell's theorem, we show here that this is irrelevant. Moreover we show that Hess and Philipp's hidden variables model (the HP model) is not local at all. Hess and Philipp do not take proper account of a crucial element in the proof of Bell's theorem: the free choice of the experimenter in the laboratory, and of the theoretician imagining a *Gedankenexperiment*, between several possible experiments. We shall convert this *freedom* into a statistical independence assumption, and show how it plays an essential role in obtaining Bell's theorem. We refer to the result of this free choice, in what follows, as the "experimental setting" or just the "setting".

Hess and Philipp's (1, 2) main criticism of Bell is his lack of consideration of time, but as we shall see below time plays an entirely trivial role in both the formulation and the proof of Bell's theorem. One may think of the correlations as determined by repeating a single measurement in many different laboratories simultaneously. In each laboratory, settings are chosen at random just once and are in force throughout that measurement.

Time naturally does play a role in real world experiments in which correlations are determined over time by repeated measurements. Memory effects may lead to dependencies and possible time variation should be accounted for. After a time delay, information from one location can arrive at another. However, by results of Gill (5), none of these effects harm experimental conclusions, *provided* the experimenter exploits his freedom to *repeatedly* randomize the settings.

Since Hess and Philipp's criticism of Bell is groundless, their purported counterexample to Bell's theorem must be in error. Indeed this is the case. Hidden deep in the elaborate construction outlined in (2) is an elementary error. To be specific, a crucial hidden variable, which correlates the measurement results at the two spatially separated ends of the experiment, is not explicitly included in the list of variables on which the model is built. Hess and Philipp take pains to show that the other hidden variables of the model can be made to satisfy "Einstein's separability", but they forget about the missing under-cover variable. We show here that this variable has a probability distribution that depends on the settings of the measuring apparatuses in both wings of the experiment. Therefore, Einstein's separability is not satisfied, i.e. the model is non-local.

We first present a proof of Bell's theorem and discuss its assumptions, emphasizing aspects of *freedom* and *local disturbances*. We conclude by explaining the error in Hess and Philipp's hidden variables model.

---
[*]Key words: Bell-type delayed choice experiment, EPR correlations, locality
[†]Mathematical Inst., Univ. Utrecht, Budapestlaan 6, 3584 CD Utrecht, Netherlands; email `gill@math.uu.nl`
[‡]Ginzton Labs, Stanford Univ., Stanford, California, 94305–4088, USA; email `gregor.weihs@stanford.edu`; and Inst. für Experimentalphysik, Univ. Vienna
[§]Inst. für Experimentalphysik, Univ. Vienna, Boltzmanngasse 5, 1090 Wien, Austria; email `anton.zeilinger@univie.ac.at`
[¶]Inst. Fizyki Teoret. i Astro., Uniwersytet Gdański, PL-80-952 Gdańsk, Poland; email: `fizmz@univ.gda.pl`

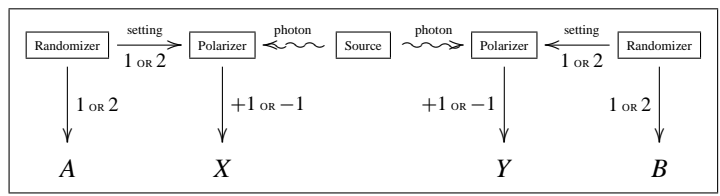

Figure 1: One trial of a Bell-type experiment.

# Freedom and Local Disturbances in Bell's Theorem

Figure 1 gives a schematic view of one trial (one pair of photons) in a Bell-type experiment, in particular, as in the experiment of Weihs et al. (6), who for the first time fully implemented Bell's requirement "the filter settings are chosen during the flight of the photons". See (4), Figure 7, p. 151. We will use the words "photons", "polarizer" and so on, but of course the picture could be applied to many different physical realizations of the Bell singlet or other suitably entangled state. In the two wings of the experiment, a "setting", denoted as $A$ and $B$, is freely and independently chosen by the local observers, and fed into the local polarizer. In each wing, it is enough to have two possible settings. We shall give them the labels 1, 2, i.e. $A$ can take the value 1 or 2 and likewise $B$. A convenient realization of that free and independent choice is for $A$ and $B$ to be outputs of two independent randomizers. The setting in each wing is fed into the measurement device (the polarizer), just before a quantum particle arrives from a distant source at the device. The measurement results in an outcome $\pm 1$. The whole process from random choice to registering the outcome takes less time than it takes a light signal to traverse from one wing of the experiment to the other. Thus the setting used in the left wing of the experiment is not available in the right wing till after the outcome there is settled, and vice versa. We will denote the outcome left by $X$ and right by $Y$; these are variables taking values $\pm 1$.

So far we have just introduced notation for four random variables $A$, $B$, $X$, $Y$ which actually get observed when the experiment (just one trial) is carried out. We next make some assumptions concerning these variables. We describe what is meant by *local realism* and show how this implies a relationship between various probabilities concerning these observable variables.

The first element of local realism is *realism* itself. By this we mean any mathematical-physical model, or a scientific standpoint, which allows one to introduce a further *eight* variables into the the model so far, which we denote by $X_{ij}, Y_{ij}$, where $i, j = 1, 2$, and which are such that

$$X \equiv X_{AB}, \quad Y \equiv Y_{AB} \quad \text{(realism)}. \quad [1]$$

In words: one may conceive, as a thought experiment or as part of a mathematical model, of "what the measurement outcomes would be, under any of the possible measurement settings"; you get to see the outcomes corresponding to the actually selected settings. No hidden variables appear anywhere in our argument beyond these eight. Given a (possibly stochastic) hidden variables theory, for instance that of (2), one will be able to define our eight variables as (possibly random) functions of the variables in that theory. According to the hypothesis of realism, in each individual experiment these variables coexist independently of which experiment is actually performed on either side.

Next, we introduce *locality*. The following is supposed to hold for all $i, j$:

$$X_{i1} \equiv X_{i2}, \quad Y_{1j} \equiv Y_{2j} \quad \text{(locality)}. \quad [2]$$

That is to say, the outcome which you would see left, under either setting, does not depend on which setting might be chosen right, and vice versa. Working under the locality assumption we write

$$X_i \equiv X_{ij}, \quad Y_j \equiv Y_{ij} \quad \text{for each } i, j. \quad [3]$$

Recall that $X$ and $Y$ are the actual outcomes corresponding to the actual settings $A$ and $B$. Under realism and locality we therefore have

$$X \equiv X_A, \quad Y \equiv Y_B. \quad [4]$$

Contained in the above is an assumption of *local disturbances*. When Alice sends her chosen label $A = i$ to her measurement device and Bob sends his chosen label $B = j$ to his, they will likely cause some further unintended disturbance. Implicit in the above is the assumption that any disturbance on the left, as far as it influences the outcome left, is not related to the coin toss nor to the potential outcomes on the right, and vice versa.

Realism and locality can be made artificially true by a mathematical trick: simply define all $X_{ij} \equiv X$, and $Y_{ij} \equiv Y$! However, this unphysical choice is ruled out by the next assumption, which we call *freedom*, often only tacit in presentations of Bell's theorem: the generation of the local settings $A$ and $B$, by randomizers or by free choice of the observers, is physically independent of the system consisting of the source and the two detectors combined which produces the *potential* outcomes $X_1, X_2, Y_1, Y_2$. The actually observed outcomes can be therefore treated as merely selected by independent setting choices as $X \equiv X_A, Y \equiv Y_B$.

Let us elaborate a bit more on the freedom assumption. Consider the following rather complicated procedure of choosing the local settings: in the left wing of the experiment, Alice shuffles and cuts a pack of cards, and decides on the basis of the chosen card (red or black) how to encode a subsequent coin toss (H or T) as setting 1 or 2, or vice versa. She selects a coin from her purse and tosses it, using the encoding just determined in order to feed either a 1 or a 2 into the communication line to 'her' polarizer. Far away, and simultaneously, Bob follows a similar procedure. Better still, he uses other randomization devices such as a roulette wheel or dartboard (he is a very poor darts player), or pseudo-random number generator with seed chosen by tossing dice, or some other physical random number generator. Note the freedom which the two persons have in: how many times to shuffle their pack of cards, which coin to pick from their purse, and so on. And notice also the complexity of the path, which, even if one believes it is essentially deterministic, results in the choice 1 or 2, to be fed into each polarizer. Finally, please note the complete independence of the above procedure from the workings of the *source and detectors* of the correlated pairs of photons. We will assume that the complete procedure used to generate $A$ and $B$ may be mathematically modeled as *independent, fair coin tosses*, thus: each of the four possible values 11, 12, 21, 22 of the pair $AB$ are equally likely. In (5) the randomizers were systems based on quantum optics.

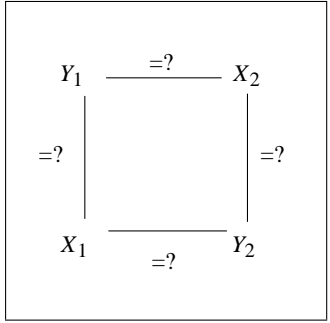

Figure 2: The number of equalities is even.

We now procede to prove Bell's inequality, and from this, Bell's theorem. The first step is to note a logical fact: arrange the four binary variables $X_1$, $Y_1$, $X_2$, $Y_2$ at the corners of a square, in the sequence just given. Each of the four sides of the square connects one of the $X_i$ with one of the $Y_j$. Now for each pair $X_i, Y_j$ ask the question: do these variables take the same value, or are they different; see Figure 2. One sees immediately that any three equalities imply the fourth; and also that any three inequalities imply the fourth. It follows that *the number of equalities always equals 0, 2 or 4*. For an algebraic proof of this fact, note that the value of $X_i Y_j$ encodes the equality or inequality of the variables $X_i$ and $Y_j$, while $(X_1 Y_2) = (X_1 Y_1)(X_2 Y_1)(X_2 Y_2)$.

Next, define $\mathbb{1}\{\ldots\}$ as the indicator variable of a specified event; that is to say, the the random variable which takes the value 1 if the event happens, 0 if not. Consider

$$\mathbb{1}\{X_1 = Y_2\} - \mathbb{1}\{X_1 = Y_1\} \\ - \mathbb{1}\{X_2 = Y_1\} - \mathbb{1}\{X_2 = Y_2\} = \Delta. \quad [5]$$

By what we have just said, $\Delta$ can only take on the values 0 and $-2$, hence its expected value is not greater than 0. Now, the expected value of a linear combination of random variables equals the same linear combination of the expected values of each variable separately. Moreover, the expected value of an indicator variable (which only takes the values 0 and 1) is equal to the *probability* of the value 1, thus equals the probability of the event in question. Therefore

$$\Pr\{X_1 = Y_2\} - \Pr\{X_1 = Y_1\} \\ - \Pr\{X_2 = Y_1\} - \Pr\{X_2 = Y_2\} = E(\Delta) \leq 0. \quad [6]$$

If the randomization procedures used by Alice and Bob are of high quality and the physical processes involved are in a physical sense independent of the workings of the source and the local detetctors, then the hidden variables $(X_1, X_2, Y_1, Y_2)$ are statistically independent of $(A, B)$. Putting it formally

$(A, B)$ is stat. indep. $(X_1, X_2, Y_1, Y_2)$ (freedom). [7]

where "stat. indep." is an abbreviation for *statistically independent of*. The technical meaning of statistical independence of $(X_1, X_2, Y_1, Y_2)$ from $(A, B)$ is that in repetitions of the experiment the relative frequencies with which the quadruple $(X_1, X_2, Y_1, Y_2)$ takes on any of its $2^4$ possible values, remain the same within each subensemble defined by each of the four possible values of the pair $(A, B)$. Statistical independence can also be called "complete lack of correlation". The notion is symmetric: the left and right hand sides of Eq. [7] can be interchanged.

Now consider the conditional probability $\Pr\{X = Y \mid A = i, B = j\}$ that the outcomes left and right are equal, given any pair of measurement settings $i$ and $j$. We write this for short as $\Pr\{X = Y \mid AB = ij\}$. By *local realism*, this equals $\Pr\{X_i = Y_j \mid AB = ij\}$. But by *statistical independence* this conditional probability is the same as the unconditional probability $\Pr\{X_i = Y_j\}$. Therefore we obtain Bell's inequality:

$$\Pr\{X = Y \mid AB = 12\} - \Pr\{X = Y \mid AB = 11\} \\ - \Pr\{X = Y \mid AB = 21\} - \Pr\{X = Y \mid AB = 22\} \leq 0. \quad [8]$$

But quantum mechanics makes the prediction, for the familiar choice of state and polarizer settings, that the expression on the left hand side of this inequality equals $\sqrt{2} - 1 \gg 0$. Hence Bell's theorem: if quantum mechanics holds, local realism is untenable.

## Hess and Philipp's Objections to Bell

Having presented Bell's theorem in this compact form we proceed to discuss Hess and Philipp's difficulties with it.

First of all, the only independence we needed was between the chosen polarizer settings, on the one hand, and the physical system of polarizers and source, on the other. Under a local realistic model, any kinds of dependencies of any hidden variables within each of the two measurement devices is allowed, as is any dependence of either one separately on the source.

Secondly, we did not mention time in our derivation at all because it was completely irrelevant. Our derivation concerned each run of the experiment. We did not compare *actual* outcomes under different settings at *different* times, but *potential* outcomes under different settings at the *same* time. Therefore, the argument in (1) Eqs **[8]** and **[9]**, or in (2), end of the paragraph following Eq. **[11]**, is completely beside the point.

To emphasize this point, consider (as a thought experiment) repeating the measurement procedure just described, not as a sequence of successive repetitions at the same locations, but in a million laboratories all over the galaxy. The prediction of local realism is that when we collect the one million sets of observed quadruples $(A, B, X, Y)$ together and compute four relative frequencies estimating the four conditional probabilities $\Pr\{X = Y \mid AB = ij\}$, they will satisfy (up to statistical error) Bell's inequality. It is of no importance that the distribution of hidden variables at different locations of the experiment might vary.

Hess and Philipp make a large number of criticisms of the assumptions of Bell's theorem with the main theme being that variables at both locations can vary in time in a dependent way, leading to dependence between the outcomes, which Bell supposedly did not take account of. Before turning to their model, we confront our formalization of the metaphysical assumptions of local realism with the idea of time variation.

How could time variation invalidate the freedom assumption? One would have to argue that because of systematic long-time periodicities in the various component physical systems concerned, the outcomes of a complex series of events involving a card shuffle, a coin toss and the free will of an experimenter at one location are interdependent and highly correlated with the potential outcome of a certain polarization measurement at a distant location. A good experimental design, with rigorous randomization of the choice of settings, makes this totally implausible.

There is an equally implausible way in which time variation could invalidate the locality assumption, connected to what we called *local disturbances* above. When we select a "1" or a "2" on a measurement device, by pressing the appropriate button, we have supposed that *only* our choice has an impact on the subsequent physics at this location. However it is clear that at the same time we will be introducing a small but uncontrolled disturbance alongside the intended binary input. In conceptual repetitions of the experiment, the length of time our finger presses the button, how hard we press the button, when precisely we do it, and so on, will vary, and each time a different disturbance is introduced into the measurement device. Could it be that this disturbance actually carries with it information about the setting being chosen in the far wing of the experiment? Well, perhaps there is a physics in which the outcomes of coin tosses, polarization measurements, and whether or not a physicist gets funding for his experiment, are determined long in advance of the events, and are encoded in minute variations in timing and pressure, so that the setting being generated by Bob is in fact already "known" at Alice's location and is unwittingly introduced by her into her apparatus along with her own coin toss (spooky, indeed). Then the outcome left, under different hypothetical settings right, could differ, and our locality assumption would fail, even though the statistical independence between the "nominal" instructions $A$, $B$ and the "hidden variables" $X_{ij}$, $Y_{ij}$ still held.

## HP model a Counterexample to Bell?

Hess and Philipp (2) construct an elaborate hidden variables model for the EPR correlations. As we will argue below, in contradiction to their claim, the HP model is nonlocal. They make a number of errors, all centering around a variable $m$ which they treat as a mere (time)-index and the related variable $i$ which they forget to treat as one of the hidden variables of the model (which is why it is missing in all their formulas except Eq. **[26]**). As we shall show, the under-cover hidden variable $i$, shared at the two measurement stations of the Bell experiment, has a probability distribution depending on the local polarizer orientations at both stations, and therefore is non-local!

An impatient reader may easily notice the inconsistency concerning the variable $m$ in the logic of the HP model, by carefully reading the five lines above Eq. **[35]** of (2). Here, $A$ and $B$ are the functions which determine the outcomes in the left and right wings of the experiment from the settings **a** and **b** and from various hidden variables and functions thereof:

> "Label the corresponding functions $A$ and $B$ as $A_{(m)}$ and $B_{(m)}$ and consider the index $(m)$ a function of the source parameter $\lambda = (\lambda^1, \lambda^2)$ and the time operators $O^1_{\mathbf{a},t}, O^2_{\mathbf{b},t}$. Then the functions $A_{(m)}$ and $B_{(m)}$ can be considered as functions of $\mathbf{a}, \lambda, \Lambda^1_{\mathbf{a},t}$, and $\mathbf{b}, \lambda, \Lambda^2_{\mathbf{b},t}$, respectively."

The first sentence introduces the variable $A_{(m)}$, with $m$ being a function of $\lambda$ and "time operators" $O^1_{\mathbf{a},t}$ and $O^2_{\mathbf{b},t}$. This means that the outcome $A$ depends, among other parameters, on $\lambda$, **a** and **b** where **b** is the setting in the other wing of the experiment! However, in the next sentence the authors write "$A_{(m)}$ can be considered as function of **a**, $\lambda$, $\Lambda^1_{\mathbf{a},t}$". They do not notice that $A_{(m)}$ also depends on $m$ and $m$ depends on **a** *and* **b**. Hess and Philipp are confused about the variable $m$ which on the one hand is treated merely as an index, determining the precise time point of the measurement, while in the mathematics it plays a less innocent role.

For the patient reader, we now go into the mathematics of the HP model in more detail. The central role in the model building is played by the Theorem on page 14231, Eqs. **[15–19]** of (2). The reader should have this paper close to hand in the following discussion. We shall show that the main formula of the Theorem, Eq. **[18]**, should be rewritten, to reveal all relevant hidden variables, in the following form

$$E\{A_{\mathbf{a}}B_{\mathbf{b}}\} = \sum_{i=-2}^{3n} \int_\Omega A_{\mathbf{a}}(u)B_{\mathbf{b}}(v)\rho_{\mathbf{ab}}(u,v;i)\mathrm{d}u\mathrm{d}v \quad \textbf{[9]}$$
$$= -\mathbf{a}\cdot\mathbf{b}.$$

The reader will find definitions of all these objects in (2), however, for the experienced reader, the notation and physical meaning of the entries to the formula should be evident. The proof of the theorem goes smoothly up to Equations **[26–27]**. Here, a quick glance tells a mathematically experienced reader that there is something missing. The definition **[26]** should be replaced by

$$\kappa(u,v;i,j) = \delta_{ij}\cdot\mathbb{1}\{i-1 \le u < i\}\cdot\mathbb{1}\{j-1 \le v < j\} \quad \textbf{[10]}$$

simply because, following the arguments of Hess and Philipp, the function $\kappa$ *does* depend on $i$ and $j$. Furthermore, one immediately notices that because of the Kronecker delta an even better choice would be

$$\kappa(u,v;i) = \mathbb{1}\{i-1 \le u < i\}\cdot\mathbb{1}\{i-1 \le v < i\}. \quad \textbf{[11]}$$

Therefore the next formula **[27]** defining the density $\rho_{\mathbf{ab}}$ should be reformulated as

$$\rho_{\mathbf{ab}}(u,v;i) = \sigma_{\mathbf{a}}(u)\tau_{\mathbf{b}}(v)\kappa(u,v;i). \quad \textbf{[12]}$$

Since the role of the function $\kappa$ is to partition the domains of the local weight functions $\sigma$ and $\tau$, it is more convenient to define new weights $\sigma_{\mathbf{a}}(u,i) = \sigma_{\mathbf{a}}(u)\cdot\mathbb{1}\{i-1 \le u < i\}$ and $\tau_{\mathbf{b}}(v,i) = \tau_{\mathbf{b}}(v)\cdot\mathbb{1}\{i-1 \le v < i\}$, so that

$$\rho_{\mathbf{ab}}(u,v;i) = \sigma_{\mathbf{a}}(u,i)\tau_{\mathbf{b}}(v,i). \quad \textbf{[13]}$$

This formula fully reveals the true nature of the distribution $\rho_{\mathbf{ab}}$. The hidden variables $u$ and $v$ are local ones, whereas the very important variable $i$ is at *both locations*, and as such it has *correlating* power.

Now we come to the crucial point: the probability distribution of the variable $i$ depends on both **a** and **b**, therefore it is of non-local nature. For simplicity, let us compute using the elaborate formulas **[24–25]** of (2) the marginal distribution of $i$, for the fixed value of $i = -2$, that is

$$\Pr(i=-2) = \int_\Omega A_{\mathbf{a}}(u)B_{\mathbf{b}}(v)\rho_{\mathbf{ab}}(u,v;i=-2)\mathrm{d}u\mathrm{d}v.$$
**[14]**

For $i = -2$ the domains of $\sigma_{\mathbf{a}}(u,i)$ and $\tau_{\mathbf{b}}(v,i)$ are $-3 \le u < -2$ and $-3 \le v < -2$ respectively, and therefore only the top formulas in **[24]** and **[25]** are relevant, giving

$$\Pr(i=-2) = |a_1|\cdot|b_1|, \quad \textbf{[15]}$$

i.e., the distribution is overtly non-local!

The subsequent elaborate transformations of the model of the Theorem, which are made with the use of the additional variable $m$, are performed by Hess and Philipp to arrange that the marginal density of the local variables $u$ and $v$ can be made uniform (unnumbered formula below Eq. **[36]**). This is all irrelevant. The correlating undercover hidden variable $i$, put under the rug by notational carelessness, has a non-local distribution. This distribution cannot be made local by the introduction of the additional variable $m$, *because the non-local character of the construction in the Theorem is retained by the final construction in the case $m = 1$!* Therefore, at least for this value of $m$, the final full model is non-local. One can show this also for the other values of $m$, but for what purpose? The model is already dead.

## Conclusions

Time is not an issue in the proof of Bell's theorem. What is crucial is the freedom of the experimenter to choose either of two settings at the same time. Hess and Philipp's hidden variables model is non-local.

**Acknowledgements** M.Ż. acknowledges KBN grant No. 5 P03B 088 20, and the Austrian-Polish programme *Quantum Communication and Quantum Information (2002–2003)*. This research was also supported by the Austrian Science Fund FWF under grant SFB 65. We are grateful to Wayne C. Myrvold whose paper (7) helped us understand the HP construction.